# Spatial Correlation Diagnostics for Atoms in Optical Lattices

John P. Grondalski[*,†], Paul M. Alsing[†], and
Ivan H. Deutsch[*]

*Department of Physics and Astronomy[*], Albuquerque High Performance
Computing Center[†], University of New Mexico, Albuquerque, NM*

jcat@unm.edu

**Abstract:**
We explore the use of first and second order same-time atomic spatial correlation functions as a diagnostic for probing the small scale spatial structure of atomic samples trapped in optical lattices. Assuming an ensemble of equivalent atoms, properties of the local wave function at a given lattice site can be measured using same-position first-order correlations. Statistics of atomic distributions over the lattice can be measured via two-point correlations, generally requiring the averaging of multiple realizations of statistically similar but distinct realizations in order to obtain sufficient signal to noise. Whereas two-point first order correlations are fragile due to phase fluctuations from shot-to-shot in the ensemble, second order correlations are robust. We perform numerical simulations to demonstrate these diagnostic tools.

**References and links**

1. P. S. Jessen and I. H. Deutsch, "Optical Lattices," Adv. At. Mol. Opt. Phys. **37**, 95-138 (1996).
2. S. Lukman Winoto, Marshall T. DePue, Nathan E. Bramall, and David S. Weiss, "Laser cooling at high density in deep far-detuned optical lattices," Phys. Rev. A **59**, R19-R22 (1999).
3. B. P. Anderson and M. A. Kasevich, "Macroscopic Quantum Interference from Atomic Tunnel Arrays," Science **282**, 1686-1689 (1998).
4. D. Jaksch, C. Bruder, J. I. Cirac, C. W. Gardiner, and P. Zoller, "Cold Bosonic Atoms in Optical Lattices," Phys. Rev. Lett. **81**, 3108-3111 (1998).
5. Dai-Il Choi and Qian Niu, "Bose-Einstein Condensates in an Optical Lattice," Phys. Rev. Lett. **82**, 2022-2025 (1999).
6. Kirstine Berg-Sørenson and Klaus Mølmer, "Bose-Einstein condensates in spatially periodic potentials," Phys. Rev. A **58**, 1480-1484 (1998).
7. E. V. Goldstein, P. Pax, and P. Meystre, "Dipole-dipole in three-dimensional optical lattices," Phys. Rev. A **53**, 2604-2615 (1996).
8. M. H. Anderson, J. R. Ensher, M. R. Matthews, C. E. Wieman, and E. A. Cornell, "Observation of Bose-Einstein Condensation in a Dilute Atomic Vapor," Science **269**, 198-201 (1995).
9. J. E. Thomas and L. J. Wang, "Quantum theory of correlated-atomic-position measurements by resonance imaging," Phys. Rev. A **49**, 558-569 (1994).
10. K. B. Davis, M. O. Mewes, M. R. Andrews, N. J. van Druten, D. M. Kurn, and W. Ketterle, "Bose-Einstein Condensation in a Gas of Sodium Atoms," Phys. Rev. Lett. **75**, 3969-3973 (1995).
11. M. R. Andrews, M.-O. Mewes, N. J. van Druten, D. S. Durfee, D. M. Kurn, W. Ketterle, "Direct, Nondestructive Observation of a Bose Condensate," Science **273**, 84-87 (1996)
12. M. R. Andrews, C. G. Townsend, H.-J. Miesner, D. S. Durfee, D. M. Kurn, W. Ketterle, "Observation of Interference Between Two Bose Condensates," Science **275**, 637-641 (1997).
13. G. Birkl, M. Gatzke, I. H. Deutsch, S. L. Rolston, and W. D. Phillips, "Bragg Scattering from Atoms in Optical Lattices," Phys. Rev. Lett. **75**, 2823-2826 (1995).
14. Hideyuki Kunugita, Tetsuya Ido, and Fujio Shimizu, "Ionizing Collisional Rate of Metastable Rare-Gas Atoms in an Optical Lattice," Phys. Rev. Lett. **79**, 621-624 (1997).

15. C. Orzel, M. Walhout, U. Sterr, P. S. Julienne, and S. L. Rolston, "Spin polarization and quantum-statistical effects in ultracold ionizing collisions," Phys. Rev. A **59**, 1926-1935 (1999).
16. P. D. Lett, R. N. Watts, C. I. Westbrook, W. D. Phillips, P. L. Gould, and H. J. Metcalf, "Observation of atoms laser cooled below the Doppler limit," Phys. Rev. Lett. **61**, 169-172 (1988).
17. Joseph W. Goodman, *Statistical Optics* (John Wiley & Sons, New York, 1985).
18. Benjamin Chu, *Laser Light Scattering, Second Edition* (Academic Press, San Diego, 1991).
19. Masami Yasuda and Fujio Shimizu, "Observation of Two-Atom Correlation of an Ultracold Neon Atomic Beam," Phys. Rev. Lett. **77**, 3090-3093 (1996).
20. M. Henny, S. Oberholzer, C. Strunk, T. Heinzel, K. Ensslin, M. Holland, C. Schönenberger, "The Fermionic Hanbury Brown and Twiss Experiment," Science **284**, 296-298 (1999).
21. William D. Oliver, Jungsang Kim, Robert C. Liu, Yoshihisa Yamamoto, "Hanbury Brown and Twiss-Type Experiment with Electrons," Science **284**, 299-301 (1999).
22. S. E. Hamann, D. L. Haycock, G. Klose, P. H. Pax, I. H. Deutsch, and P. S. Jessen, "Resolved-Sideband Raman Cooling to the Ground State of an Optical Lattice," Phys. Rev. Lett. **80**, 4149-4152 (1998).
23. Ivan H. Deutsch and Poul S. Jessen, "Quantum-state control in optical lattices," Phys. Rev. A **57**, 1972-1986 (1998).
24. B. Saubaméa, T. W. Hijmans, S. Kulin, E. Rasel, E. Peik, M. Leduc, and C. Cohen-Tannoudji, "Direct Measurement of the Spatial Correlation Function of Ultracold Atoms," Phys. Rev. Lett. **79**, 3146-3149 (1997).
25. E. V. Goldstein, O. Zobay, and P. Meystre, "Coherence of atom matter-wave fields," Phys. Rev. A **58**, 2373-2384 (1998).
26. Rodney Loudon, *The Quantum Theory of Light, Second Edition* (Oxford University Press, New York, 1983).

## 1. Introduction

Optical lattices, periodic arrays of microscopic potentials induced by the ac Stark effect of interfering laser beams, can be used to trap ultra-cold atoms [1]. Experiments with these lattices are achieving ever higher atomic densities through special cooling techniques [2] and by loading Bose-Einstein condensates (BECs) into lattices [3]. At such densities atom-atom interactions play an important role in the dynamics [4-7], and may lead to the formation of small scale structure within the lattice. Spatial distributions of cold atomic gases are usually probed directly by absorption spectroscopy [8], near-resonance florescence spectroscopy [9], or off-resonance spectroscopy [10,11]. The resolution of these imaging techniques is fundamentally limited by the wavelength of the external probe laser. In addition, the photons used to probe the atomic distribution impart on average an energy of $(\hbar k)^2/(2M)$ ($\hbar k$ is the photon momentum and $M$ is the mass of the atom). For cold atomic samples, these "recoil kicks" generally heat the sample very quickly, although in the case of off-resonance spectroscopy this heating has been suppressed by a factor of order 100, allowing for multiple images before the sample is destroyed [11]. These imaging techniques typically integrate the signal along one dimension of the atomic cloud thus measuring column densities. Some experiments have achieved three-dimensional resolution through tomographic methods [12]. Spatial information can also be inferred by measuring Bragg reflection from an atomic sample [13] or atomic collision rates [14,15].

   In this article we consider Time of Flight (TOF) diagnostics [16] whereby atoms initially trapped in an optical lattice are released and allowed to freely expand for a time $t$, after which they are counted with point-like detectors, such as a microchannel plate array (MCP) in the case of meta-stable noble gas atoms [14,15]. Recoil heating is eliminated because no external fields are used and additional restrictions on the resolution imposed by a probe laser no longer apply. We consider same-time spatial correlations of order one and two in the detection plane and find Fourier relations between these functions and the initial atomic distribution. These relations are completely separable in three dimensions allowing for the possibility of 3D resolution.

In order to achieve a sufficient signal-to-noise ratio in any practical experiment, an ensemble average of many TOF measurements will be necessary. When one does this, the spatial information contained in the first order correlation function is generally washed out due to random phase fluctuations in the shot-to-shot realizations. By contrast, the second order correlation function is more robust because it is insensitive to this phase. An analogous situation exists in optics. The Michelson stellar interferometer which measures first-order field correlations is very sensitive to atmospheric fluctuations whereas the Hanbury-Brown Twiss interferometer which measures intensity correlations is more stable [17].

The remainder of this article is organized as follows. In Sec. 2.1 we review the relationship between TOF measurements and the wave function of a trapped atom. We show that for an atomic sample of equivalent atoms, properties of the local wave function can be deduced. We describe a nontrivial example whereby this technique can be used to measure the wave function of an individual atom, and hence, a possible way to study quantum coherence. In the next two sections we examine ways to measure statistical properties of the distribution function for atoms throughout the lattice (i. e. the small scale structure) with two point correlation functions. In Sec. 2.2 we consider a Young's double slit experiment and establish the relationship between the complex coherence factor and the spatial distribution of the atomic sample. We show that the mean position of the atom is mapped onto the phase of the complex coherence factor, a quantity susceptible to shot-to-shot fluctuations in the sample. In Sec. 2.3 we consider the possibility of deducing spatial information from the next higher correlation function. Second order correlations of scattered laser light have been used successfully to measure the size and shape of macromolecules in solution as well as in colloidal suspensions [18]. In contrast to the photon coincidences measured in these experiments, we propose the use of atom coincidence counts, similar to recent experiments that employed time-delayed coincidence counting to analyze degenerate Bose [19] and Fermi [20,21] gases. There exists a Fourier relationship between the second order correlation function at the detector plane and the probability for a pair of atoms to have a certain separation, independent of the pair's absolute location in the lattice. In Sec. 3 we present the results of numerical simulations of ensemble averaged TOF experiments. Averaging washes out information about the mean positions of atoms contained in the first order correlation function but important information is retained in second order. Finally, in Sec. 4 we summarize our results.

## 2. Spatial Correlation Functions

### 2.1 Atomic Density Measurements

The normalized first order correlation function at a single position, $g^{(1)}(x,x)$, is a measure of the atomic density $n(x)$. Such correlations can be measured using the well known time of flight (TOF) technique in which a trapped atomic sample is released from an optical lattice and allowed to fall onto a detection plane. The arrival times of the atoms are measured and the initial momentum distribution of the atoms is inferred [16]. Consider, for example, an atom initially localized at lattice site $i$ in the state, $\Phi(x' - x'_i, t' = 0)$, where $\Phi(x')$ is the local wave function, with primed variables $(x', t')$ denoting the space-time coordinates of the optical lattice plane and the unprimed variables $(x, t)$ denoting the detection plane. An atom allowed to freely expand for a sufficiently long time after the trapping lasers are turned off can be considered to be in the far field and the wave function at a later time, $\Psi(x, t)$, is related to the initial wave function through a Fourier transform,

$$\Psi(x,t) = \frac{\sqrt{-i}}{L} \exp\left[i\frac{2\pi}{L^2}\left(\frac{x^2}{2} + x\, x'_i\right)\right] \mathcal{F}[\Phi(x')]_{u=\frac{x}{L^2}}. \quad (1)$$

Here $u$ is the reciprocal coordinate and $L$ is an *effective* length given by, $L^2 = h(t-t')/M$, which is a measure of the time evolved between the initial and final state. The mean location of the initial wave packet, $x'_i$, is mapped onto the phase of the final wave function as a consequence of the well known shift theorem from Fourier analysis. This implies that in the far field, the detected signal is proportional to the absolute value squared of the momentum space wave function and is insensitive to the initial position of the atomic wave packet. For a large collection of incoherent atoms, each in an arbitrary mixed state, the measured TOF signal is given by a statistical average,

$$n(x) = \sum_{j=1}^{R} p_j \, \frac{1}{L^2} \, |\mathcal{F}[\Phi_j(x')]_{u=\frac{x}{L^2}}|^2, \quad (2)$$

where $p_j$ is the classical probability for the $j^{th}$ local wave function to occur. If the atomic sample consists of $R$ localized atoms in quantum mechanical pure states which differ only by translation to a given lattice site, as recently demonstrated by Hamman *et al.* [22], then one can partially reconstruct the initial wave function of the individual atoms from the Wiener-Khintchine theorem subject to the usual limitations imposed by the loss of phase information caused by taking the modulus [17]. This somewhat surprising result can be made intuitive if one considers an optical analogy. Given light illuminating a small random cluster of identical pinholes, as long as the photodetector is in the far field and the distance between the pinholes is not too large, the intensity will simply be the diffraction pattern of a single pinhole resulting from the the incoherent sum of the individual, completely overlapping, diffraction patterns.

Non-trivial dynamical information about atoms trapped in optical lattices can be deduced from the results above. Suppose one arranges a 1D lattice of $N$ double wells using a configuration of counterpropagating lasers with wavelength $\lambda$ whose linear polarization have a relative angle $\theta$ [23]. The $i^{th}$ double well is located at $x'_i$ and the well separation is set by $\Delta\xi' = (\lambda/(2\pi))\tan^{-1}(\tan(\theta)/2)$. With the appropriate cooling and preparation of the initial state, the wave packet dynamics is essentially restricted to a two dimensional Hilbert space spanned by two macroscopically separated Gaussians $\Phi_0$ of width, $\sigma'$, centered about, $\pm\Delta\xi'/2$. A general state is given by the wave packet,

$$\Psi(x') = c_1 \, \Phi_0(x' - \Delta\xi'/2) + e^{i\phi} \, c_2 \, \Phi_0(x' + \Delta\xi'/2) \quad (3)$$

Here, $c_1$ and $c_2$ are the (real) probability amplitudes and $\phi$ is the relative phase between the Gaussians. If this wave packet is allowed to freely expand, the atomic density at the detector plane will have a Gaussian envelope with fringes whose spacing is given by, $d_f = L^2/\Delta\xi'$, (See Fig. 1),

$$|\Psi(x)|^2 = \sqrt{\frac{1}{2\pi\sigma^2}} \, \exp\left(-2\left(\frac{x}{2\sigma}\right)^2\right) \left(1 + 2c_1c_2 \cos\left(2\pi\, \frac{x\Delta\xi'}{L^2} + \phi\right)\right). \quad (4)$$

Here $\sigma = L^2/(4\pi\sigma')$ is the width of the fringe envelope in the detection plane. Observation of the fringes requires that the size of the sample, $S$, be much smaller than size of the fringes, $S \ll d_f$ (i. e. we are sufficiently in the far field). In the absence of decoherence the complete initial wave function can be inferred. The relative phase can be deduced from the shift of the center of the diffraction pattern with respect to the fringe envelope and the probability amplitudes can be obtained from the visibility of the fringes as a function of time. In the presence of decoherence the visibility of these fringes will decay with time and will not exhibit recurrences characteristic of the macroscopic superposition state. This could be a useful diagnostic to measure the decohering effects of the lattice environment [23]. Similar interference patterns have been measured in TOF to extract the temperature of an atomic sample, cooled via velocity-selective-coherent-population-trapping (VSCPT) [24]. The effectiveness of the above diagnostic depends on one's ability to prepare identical pure states. Intensity inhomogeneities of the trapping laser or magnetic field gradients will cause wave functions at different lattice sites to vary slightly, thus broadening the signal, although this effect could be reduced through the use of apertures.

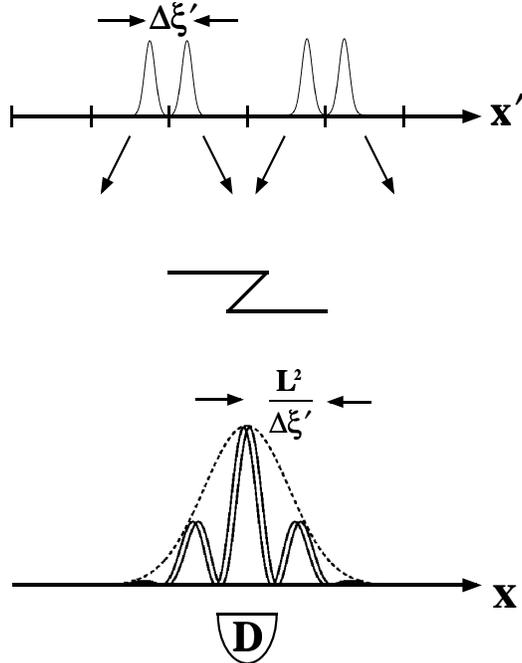

Figure 1.  Schematic of a TOF experiment. The atomic wave functions initially consists of a superposition of two Gaussians separated by $\Delta \xi'$ with relative phase $\phi = 0$ and probability amplitudes $c_1 = c_2 = 1/\sqrt{2}$. The resulting completely incoherently overlapping fringe pattern, has a fringe spacing of $L^2/\Delta \xi'$.

2.2  *Visibility Measurements*

We consider next the normalized first order correlation function, at different positions, $g^{(1)}(x_1, x_2)$, defined as the *complex degree of coherence* [17]. The modulo of this function,

$$V(\Delta x) = |g^{(1)}(\Delta x/2, -\Delta x/2)| \qquad (5)$$

corresponds to the visibility of fringes formed by a Young's double slit experiment with slit spacing $\Delta x = x_2 - x_1$. In Michelson stellar interferometry one measures the visibility of the interference pattern as a function of $\Delta x$ to deduce the spatial intensity distribution of the source [17]. We consider the atom-optic version here as a diagnostic of the distribution of atoms throughout the lattice (i.e. the small scale structure).

Suppose that a 1D lattice has $N$ sites with lattice constant $w'$. For simplicity we take the wells to be harmonic at each site and the atomic state to be thermal so that the local wave function is Gaussian. An atom in the vibrational ground state, initially located at $x'_j = jw'$ ($j = 0, \cdots, N-1$), will expand into an approximate plane wave in the far field (See Fig. 2),

$$\Psi(x,t) \propto e^{ik_j x} \text{ where } k_j = j2\pi\frac{w'}{L^2}. \tag{6}$$

The atomic field impinging on the double slit detector plane can thus be modeled as a set of $N$ discrete plane waves with a mode spacing of $\Delta k = 2\pi w'/L^2$. We define creation and annihilation operators, $\hat{a}_j^\dagger$ and $\hat{a}_j$, for the $j^{th}$ mode. The position space annihilation operator $\hat{b}_i$ at detector position $x_i$ is given by,

$$\hat{b}_i = \frac{1}{\sqrt{N}} \sum_j \hat{a}_j e^{ik_j x_i}. \tag{7}$$

We will take our atoms to be Bosons (true for most laser cooled species) with creation and annihilation operators satisfying the usual canonical commutation relations, though the analyis for first order correlation functions is identical for the case of Fermions. We further simplify our analysis by making some assumptions about the preparation of the atomic sample. In a typical laser cooling experiment, near resonant scattering and collisions prohibit two atoms from occupying the same lattice site. We therefore only consider the case where a lattice site is either empty or contains one atom. In full analogy with quantum optics we assume that the apparatus detects an atom by removing it from the field (i.e. an MCP), so that atom detection can be described by normally ordered creation and annihilation operators. We will not consider here other definitions of coherence based on other detection schemes (e.g. florescence and nonresonant imaging [25]).

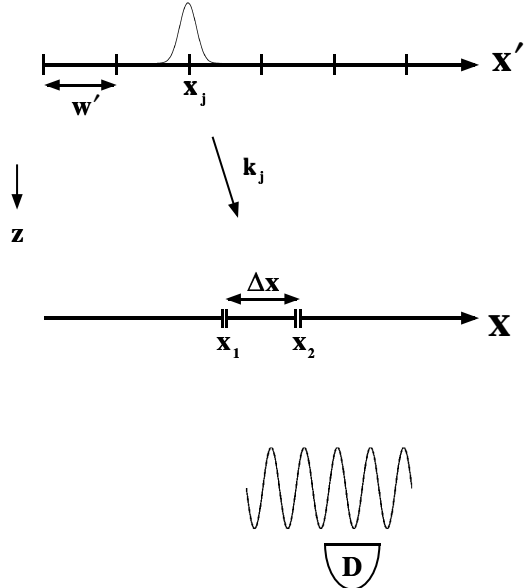

Figure 2. Schematic of an experiment that measures same-time two-point first order spatial correlations of an atomic field. By measuring the visibility of the fringes as a function of slit spacing one can deduce the atomic distribution.

The detection time can be approximated as instantaneous if it is much shorter than the coherence time, given by the length of the wave packet in the $z$-direction divided by the group velocity. Furthermore, if we assume that an equal flux impinges on each slit, then the complex degree of coherence is [26],

$$g^{(1)}(x_1, x_2) = \frac{G^{(1)}(x_1, x_2)}{\sqrt{G^{(1)}(x_1, x_1)\, G^{(1)}(x_2, x_2)}},\qquad(8)$$

$$\begin{aligned} G^{(1)}(x_1, x_2) &= \langle \hat{b}_1^\dagger \hat{b}_2 \rangle \\ &= \frac{1}{N} \sum_{j,\ell} e^{i\,(k_j x_2 - k_\ell x_1)}\, \langle \hat{a}_\ell^\dagger \hat{a}_j \rangle. \end{aligned}$$

Given an atomic field we can evaluate this expression and associate it with detection through a double slit.

For concreteness consider a single atom located at $x'_j$. The state of this system is described by the one-atom Fock state, $|0, \cdots, 1_{k_j}, \cdots, 0\rangle$, so that the complex coherence factor is given by,

$$g^{(1)}(x_1, x_2) = e^{i\,k_j(x_2 - x_1)}.\qquad(9)$$

From this expression we see that the mean position of the atom in the lattice maps onto the phase of the correlation function or the shift of the zero-delay of the fringes in a double-slit experiment. It is simple to generalize this for ensemble averages of many atomic field states,

$$g_\ell^{(1)} = \sum_{j=0}^{N-1} P_j^{(1)} e^{i\,(k_j\,\ell\Delta x)} = \sum_{j=0}^{N-1} P_j^{(1)}\, e^{i\,(j\Delta k\,\ell\Delta x)},\qquad(10)$$

where $g_\ell^{(1)} = g^{(1)}(-\frac{\ell}{2}\Delta x, \frac{\ell}{2}\Delta x)$ ($\ell = 0, \cdots, N-1$), is the complex degree of coherence for a slit spacing of $\ell\Delta x$ and $P_j^{(1)}$ is the probability for lattice site $j$ to contain an atom. The fundamental slit spacing $\Delta x$ is set by the size of the atomic sample, $(N-1)w'$; a slit spacing of $(N-1)\Delta x$ is necessary to resolve two atoms separated by a $w'$. Thus, we see that the atomic probability distribution is related to the complex degree of coherence through a discrete Fourier transform. This is the atom-optic analog of the Van Cittert-Zernike theorem [17]. The inverse relationship is,

$$P_j^{(1)} = \frac{1}{2\pi N} \sum_{\ell=0}^{N-1} g_\ell^{(1)}\, e^{-i\,(j\Delta k\,\ell\Delta x)}.\qquad(11)$$

In principle, Eqs (10)-(11) can completely determine the spatial distribution of an atomic sample from an ensemble of double slit experiments. However, unlike the example considered in Sec. 2.1 where a Young-type interference pattern was built into the initial wave function of the double well and seen in same-position first order correlations, in this case coherence is sampled at two spatially separated points masked by the double slit. This is a low flux measurement requiring ensemble averaging of multiple realizations of similarly prepared systems to acquire sufficient signal-to-noise. Furthermore, even with high atomic flux, one must repeat the measurement at a variety of different slit spacings to deduce the visibility dependence. Thus, Eqs. (10)-(11) are a useful diagnostic only if the atomic distribution can be *exactly* reproduced for each run of the experiment. In general, however, the atomic distribution will vary from shot-to-shot, causing uncontrollable phase shifts in the fringe pattern which will wash out the spatial information, even for atomic samples that are statistically similar. For this reason we consider higher order spatial correlations.

## 2.3 Coincidence Count Measurements

We now consider the atom-optic analog of Hanbury-Brown Twiss stellar interferometry which uses photon coincidence counting as a function of $\Delta x$ to deduce the spatial intensity distribution of the source [17]. The normalized same-time second order spatial correlation function at different positions, $g^{(2)}(x_1, x_2)$, corresponds to atom coincidence counts between two point-like detectors located at $x_1$ and $x_2$ in the detection plane [26] (See Fig. 3),

$$g^{(2)}(x_1, x_2) = \frac{G^{(2)}(x_1, x_2; x_2, x_1)}{G^{(1)}(x_1, x_1) \, G^{(1)}(x_2, x_2)}, \tag{12}$$

$$\begin{aligned} G^{(2)}(x_1, x_2; x_2, x_1) &= \langle \hat{b}_1^\dagger \hat{b}_2^\dagger \hat{b}_2 \hat{b}_1 \rangle \\ &= \frac{1}{N^2} \sum_{j,j',\ell,\ell'} e^{i\,((k_j - k_{j'})x_1 + (k_\ell - k_{\ell'})x_2)} \langle \hat{a}_{j'}^\dagger \hat{a}_{\ell'}^\dagger \hat{a}_\ell \hat{a}_j \rangle. \end{aligned}$$

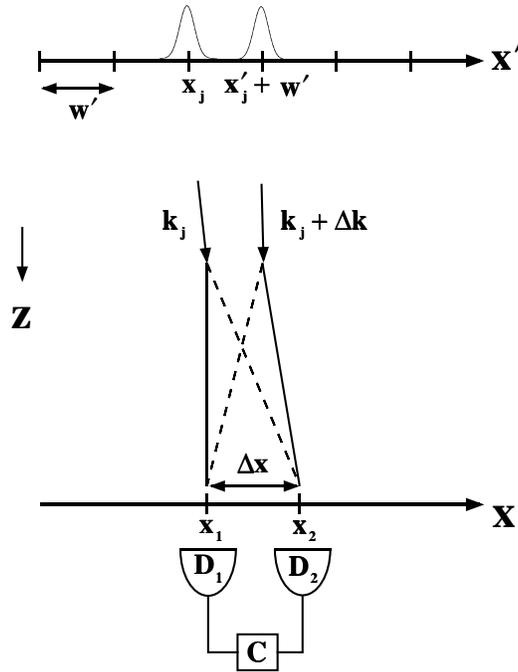

Figure 3. Schematic for an experiment that measures same-time second order spatial correlations of an atomic field. The interference arises from the two possible ways that the atoms can be jointly detected, denoted by the solid and dotted paths.

Again for concreteness consider two Bosons separated by one lattice site, $x'_j$ and $x'_j + w'$, with state vector, $|0, \cdots, 1_{k_j}, 1_{k_j + \Delta k}, \cdots, 0\rangle$, where $\Delta k = 2\pi w'/L^2$. Inserting this in Eq. (12) gives, $2g^{(2)}(x_1, x_2) - 1 = \cos(\Delta k(x_2 - x_1))$. The difference in position between the two atoms maps onto the spatial period of the cosinusoidially varying coincidence counts. For Fermions, the fringes receive a $\pi$ phase shift due to their anticommutation relations. The interference exists even with *no quantum entanglement* between the two atoms. The interference term arises from the two possible indistinguishable paths that lead to joint detection (See Fig. 3). This relation can be generalized for field states of $R$ atoms distributed throughout $N$ lattice sites. We define $g^{(2)}_\ell = g^{(2)}(-\frac{\ell}{2}\Delta x, \frac{\ell}{2}\Delta x)$ for two detectors separated by $\ell \Delta x$ ($\ell = 0, \cdots, N-1$), given the state $a_1^\dagger a_2^\dagger, \cdots, a_R^\dagger |0\rangle$ we

find
$$\frac{R^2}{R(R-1)} g_\ell^{(2)} - 1 = \sum_{j=0}^{N-1} P_j^{(2)} \cos(j\Delta k\, \ell\Delta x). \tag{13}$$

$P_j^{(2)}$ is the probability for two atoms to be separated by $j$ lattice constants $jw'$ ($j = 0, \cdots, N-1$), or equivalently, for two atomic plane waves to impinge the detector plane with a mode spacing of $j\Delta k$. We see that for $R = 2$, $g_0^{(2)} = 1$, reflecting the perfect second-order coherence for the state of exactly two atoms and $g_0^{(2)} \to 2$ as $R \to \infty$, which is the usual bunching factor associated with a highly chaotic macroscopic distribution [26]. Note, $P_j^{(2)}$ depends only on the relative mode spacing and is independent of the absolute location of a pair of atoms. Such information is mapped onto the phase of the first order correlation function as discussed in Sec. 2.2 but is irrelevant to the second order correlation function.

The sum can be analytically extended to negative values of $j$ ($j = -N, \cdots, N-1$) since $g^{(2)}$ is symmetric under reflections about zero, and we obtain the following discrete Fourier relation,

$$g_\ell^{(2)} - 1 = \sum_{j=-N}^{N-1} \frac{P_j^{(2)}}{2} e^{i\, j\Delta k\, \ell\Delta x} \tag{14}$$

and its inverse
$$\frac{P_j^{(2)}}{2} = \frac{1}{2\pi(2N)} \sum_{\ell=-N}^{N-1} (g_\ell^{(2)} - 1) e^{-i\, j\Delta k\, \ell\Delta x}. \tag{15}$$

Similar relations can be obtained for Fermions by making the transformation $(g_\ell^{(2)} - 1) \to -(g_\ell^{(2)} - 1)$.

The first and second order correlation functions are related under the assumption that the atomic positions satisfy Gaussian statistics,

$$g^{(2)}(x_1, x_2) = 1 + |g^{(1)}(x_1, x_2)|^2. \tag{16}$$

$g^{(2)}(x_1, x_2)$ can then be related to the autocorrelation of the initial atomic distribution through the Wiener-Khintchine theorem [17]

$$g^{(2)}(x_1, x_2) - 1 = \mathcal{F}[P^{(1)}(x') \otimes P^{(1)}(x')]. \tag{17}$$

We will not restrict our attention to this case in the results presented below.

To determine the practical resolution of this detection scheme, note that a pair of atoms spaced by $w'$ in the optical lattice plane will result in coincidence count period of, $\Lambda = L^2/w'$. The smallest atomic separation in the lattice plane that can be resolved is determined by the width of the Gaussian envelope which modulates the interference fringes. Given a wave packet with initial width $\sigma'$, located at $x_i'$, the wave function in the detection plane will be given by,

$$\Psi(x, t) \simeq \left(\frac{1}{2\pi\sigma^2}\right)^{\frac{1}{4}} e^{-i\pi(\frac{x'-x_i'}{L})^2} e^{-(\frac{x'-x_i'}{2\sigma})^2}, \tag{18}$$

where $\sigma$ was defined in Sec. 2.1. For the situation of cesium atomic wave packets freely expanding for $\sim 1$ sec, $w' \simeq 0.5\mu m$, and $\sigma' \simeq 30$ nm, then $\Lambda \simeq \sigma \simeq 1$ cm. Also, for any case of practical significance, the small scale information will be contained in the "wings" of the Gaussian in the detection plane where the phase varies quadratically.

The above Fourier relations will still be valid, however, if the detectors are located on the wave front of constant phase, since the Van-Cittert Zernike theorem is valid under a Fresnel approximation [17]. This can be accomplished with a curved paraboloid detection surface, symmetric coincidence counts, or a flat detection surface with appropriate electronic time delays introduced. Finally, these relations are completely separable and can be extended to three dimensions. For the $z$-direction (the direction that the atoms are falling) same-time spatial coincidences are replaced by same-position, temporal coincidences, with the transformation given by the appropriate dynamical equations. If the atoms are falling under the influence of gravity, then $x = \frac{g}{2}t^2$, and a detector spacing of $\Delta x$ corresponds to a detection delay time of $gt\Delta t$.

## 3. Results

In principle, the density operator, $\hat{\rho}$, offers a complete description of the atomic distribution. For our assumptions about the nature of the sample with no coherences between different lattice site, we need only consider diagonal matrix elements,

$$P^{(n)}(x'_1, x'_2, \cdots, x'_n) = \langle 1_{k_1}, 1_{k_2}, \cdots, 1_{k_n} | \hat{\rho} | 1_{k_1}, 1_{k_2}, \cdots, 1_{k_n} \rangle, \quad (19)$$

corresponding to the nth order joint probability detection.

In this work we will restrict our attention solely to the first and second order probabilities. We have seen that $g^{(1)}(x_1, x_2)$ allows for a direct measure of $P_j^{(1)}$, the probability for an atom to be located at $x'_j$, and $g^{(2)}(x_1, x_2)$ provides for a direct measure of the $P_j^{(2)}$, the joint probability for two atoms to be separated by $j$ lattice spacings, $w'$, independent of the absolute location of the pair,

$$P_j^{(2)} = \sum_{\ell=1}^{N} P^{(2)}(x'_\ell + jw', x'_\ell). \quad (20)$$

The spatial information contained in $P_j^{(1)}$ and $P_j^{(2)}$ is different. To see this, factor the joint probability under the sum using Bayes' rule,

$$P_j^{(2)} = \sum_{\ell=1}^{N} P_\ell^{(1)} P(x'_\ell + jw' | x'_\ell), \quad (21)$$

where $P(x'_\ell + jw' | x'_\ell)$ is the conditional probability for an atom to be located at $x'_\ell + jw'$ given that an atom is at $x'_\ell$. Only in the special case that the lattice sites are statistically independent so that $P(x'_\ell + jw' | x'_\ell) = P_{\ell+j}^{(1)}$, do we obtain $P_j^{(2)}$ from an autocorrelation of $P_j^{(1)}$,

$$P_j^{(2)} = \sum_{\ell=1}^{N} P_\ell^{(1)} P_{\ell+j}^{(1)}. \quad (22)$$

This is an example of the Wiener-Khintchine theorem, Eq. (17).

We have performed computer simulations of TOF experiments that measure the first and second order spatial correlations discussed above. The lattice had 256 lattice sites and a fill factor of $\simeq 10\%$. The simulation consisted of ensemble averaging the complex coherence factor and atom coincidence counts over 500 runs. The averaged data is inverted with the appropriate Fourier relation given above to obtain $P_j^{(1)}$ and $P_j^{(2)}$ respectively. In the first simulation, the atoms were distributed in the lattice according to a conditional "bunched" distribution in which a seed point $x'_\ell$ was picked and then the atomic distribution was conditioned on it in such a way as to cluster around it,

$$P(x'|x'_\ell) = \sqrt{\frac{1}{2\pi\tau}} e^{-2(\frac{x'-x'_\ell}{2\tau})^2}. \qquad (23)$$

In order to elucidate the effects of shot-to-shot phase fluctuations, the seed point was picked in two ways. First we fixed the seed at lattice site $N = 128$ for each run of the gedenken experiment and in the second we let the initial seed point vary randomly (See Fig. 4).

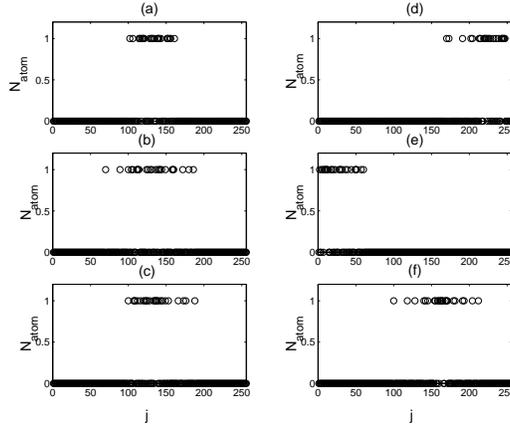

Figure 4. Atomic distribution bunched around a chosen "seed point". First column (a)-(c) shows samples with a fixed seed at site $N = 128$. The second column, (d)-(f), shows the atomic distribution for a randomly varying seed point.

The results for $P_j^{(1)}$ and $P_j^{(2)}$ given the fixed seed point are shown in Fig. 5a-b and we find that both correlation functions contain useful spatial information. When the seed point is varied randomly for each run of the experiment, so that the atoms tended to cluster in a different parts of the lattice, we see that the spatial information contained in $P_j^{(1)}$ completely washes out, while the spatial information in $P_j^{(2)}$ is unaffected (Fig. 5c-d). The first order correlations $g^{(1)}(x_1, x_2)$ depend on the absolute locations of atoms and on average it sees a randomly filled lattice. In contrast, $g^{(2)}(x_1, x_2)$ measures only the relative locations of atoms independent of the absolute location of the cluster.

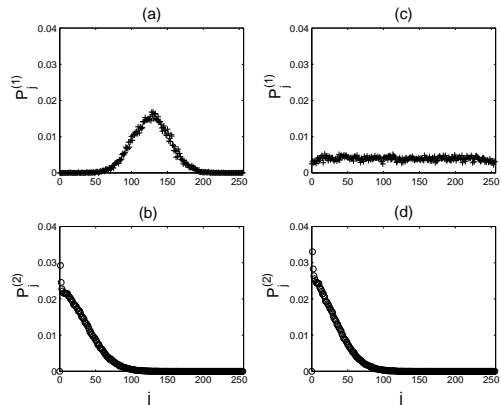

Figure 5. Probability for an atom to be located a site $j$, $P_j^{(1)}$, and for two atoms to be separated by $j$ sites, $P_j^{(2)}$, for a fixed seed point, (a) and (b), and for a randomly varying seed point, (c) and (d).

To illustrate the capability of atomic coincidence counting, we carried out simulations for various distributions: random, "bunched", "anti-bunched", macroscopic-periodic. The results are compared in Fig. 6. Note that our coincidence period was taken to be just large enough to resolve the lattice spacing. Larger detection areas would result in a "picket fence" distribution which would explicitly display the periodicity of the lattice.

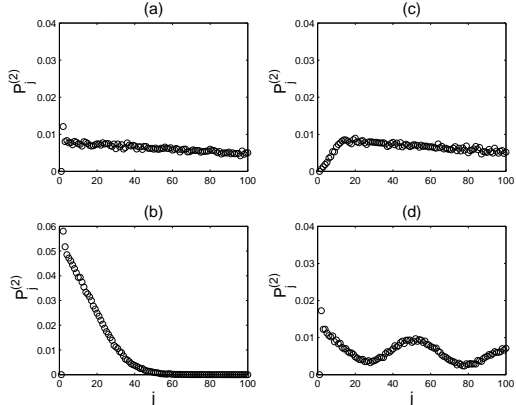

Figure 6. Probability for atoms to be separated by $j$ lattice sites, obtained by Fourier transform of simulated coincidence count measurements. Results are shown for atomic distributions which are (a) a random, (b) "bunched", (c) "anti-bunched", and (d) macroscopic variation on a "super-lattice".

An important simplification arises when the atomic sample is characterized by a conditional probability with the functional form

$$P(x'|x'_\ell) = f(x' - x'_\ell). \tag{24}$$

Upon substituting this into Eq. (21) one finds,

$$P_j^{(2)} = \sum_\ell P_\ell^{(1)} f(x'_\ell + jw' - x'_\ell) = f(jw'). \tag{25}$$

In this case of stationary statistics, the joint probability, as measured by $g^{(2)}(x_1, x_2)$, is a direct measure of the relative conditional probabilities of the lattice, independent of global properties of the lattice such as intensity or magnetic field inhomogeneities.

## 4. Summary

For very cold atoms trapped in an optical lattices the wave nature of an atom becomes manifest. We have explored the possibility of exploiting this feature to image the atomic distribution using both first and second, same-time, atomic spatial correlation functions in TOF diagnostics. One finds that information about a single atomic wave function can be inferred from the atomic density in the detection plane for a lattice filled with atoms whose quantum mechanical states are identical up to translation. The ability to measure this quantity depends on having a lattice that is homogeneous, but the SNR obtained for a single run of a TOF experiment, for a typical lattice, is large. Information about the distribution of atoms throughout the lattice can be obtained from both first and second order correlation functions at different spatial points through Fourier relations that connect the measured signal and the initial atomic distribution. The necessity of averaging many TOF experiments in order to obtain sufficient signal generally leads to fluctuations which wash out the interference fringes associated with $g^{(1)}(x_1, x_2)$. In contrast, $g^{(2)}(x_1, x_2)$ is more robust and can be a more useful diagnostic. We have performed computer simulations of these detection schemes which illustrate the salient features.

## Acknowledgements

The authors gratefully acknowledge Sudakar Prasad, Steven Rolston, Simone Kulin, and Gavin Brennen for many useful discussions. This research was supported by NSF grant 9732456 and the Albuquerque High Performance Computing Center.